\documentstyle[aps,twocolumn,psfig]{revtex}
\begin{document}
\title{Finite temperature numerical renormalization group\\
study of the  Mott-transition}
\author{R. Bulla$^a$, T.A. Costi$^b$, and D. Vollhardt$^a$}
\address{$^a$Theoretische Physik III, Elektronische Korrelationen und Magnetismus, 
Institut f\"ur Physik,\\ Universit\"at Augsburg,  D-86135 Augsburg, Germany\\
$^b$Institut Laue-Langevin,        
B.P. 156 - 38042 Grenoble Cedex 9, 
France
}
\draft
\maketitle
\begin{abstract}
Wilson's numerical renormalization group (NRG) method 
for the calculation of dynamic properties of impurity
models is generalized  to investigate
the effective impurity model  of the
dynamical mean field theory at finite temperatures.
We calculate the spectral function
and self-energy for the Hubbard model on 
a Bethe lattice with infinite coordination number
 directly on the real frequency axis and investigate
the phase diagram for the Mott-Hubbard metal-insulator transition.
While for $T<T_{\rm c}\approx 0.02W$ ($W$: bandwidth)
we find hysteresis with first-order transitions
both at $U_{\rm c1}$ (defining the insulator to metal transition)
and at $U_{\rm c2}$ (defining the  metal to insulator transition),
at $T>T_{\rm c}$ there is a smooth crossover from metallic-like to 
insulating-like solutions.

\end{abstract}
\section{Introduction\label{sec:1}}

During the past decade, the development and application of
the dynamical mean field theory (DMFT) has led
to a considerable increase in our understanding of strongly
correlated electron systems. The DMFT has originally been
derived from the limit of infinite spatial dimensionality
(or, equivalently, infinite lattice connectivity) of lattice 
fermion models, such as the Hubbard model \cite{MV}.
In this limit, the self-energy becomes purely local \cite{mh.89}, 
which is a consequence
of the required scaling of the hopping matrix element
$t\!=\!t^\ast/\sqrt{d}$, with $t^\ast$ fixed and $d$ the 
lattice dimension.

It has been realized in the work of Jarrell \cite{Jar92} and
Georges and Kotliar \cite{Geo92} that such a local self-energy
can be calculated from a much simpler, but nevertheless
highly non-trivial model: the single impurity Anderson model
(SIAM) \cite{And61}. The self-energy of the SIAM is
local because the Coulomb correlation in this model only
acts on the impurity site. The difference between
the SIAM and the lattice model under consideration is then
built in via a self-consistency condition \cite{Geo96}.
In this way, the DMFT became a powerful tool for the
investigation of various lattice models such as  the Hubbard model
and the periodic Anderson model (for a review see \cite{Geo96}).
The success of this approach, however, depends on the
availability of
reliable methods for the calculation of the self-energy of an effective
SIAM.
Perturbative methods, such as the iterated perturbation theory 
\cite{Geo96} or
the non-crossing approximation \cite{Pru93}, have been shown to give 
qualitatively correct results for a variety of physical problems.
The numerical implementation of these methods allows one to solve
the impurity model with a minimum of computational effort
(typically a few seconds on a workstation) so that the
relevant parameter space of the model can be scanned
very quickly.

However, most of the phenomena of interest in strongly
correlated systems are
inherently non-perturbative, so that none of the parameters
in the Hamiltonian can be regarded as a small perturbation.
In general we 
therefore  have to apply non-perturbative methods,
even in cases where perturbative approaches such as the 
iterated perturbation theory or
the non-crossing approximation appear to give
a complete picture of the solution.

The most widely used non-perturbative method in this
context is the quantum Monte-Carlo approach \cite{Jar92,Geo96}. 
The advantages
of this method are its flexibility (a wide range of physical problems
can be studied with only relatively minor changes in the
program) and the possibility of obtaining a `numerically exact'
solution of $G(\tau)$, the single-particle Green function on the
imaginary time axis. The main disadvantage of the 
quantum Monte-Carlo method is the drastic increase
of computation time upon either increasing the Coulomb repulsion
$U$ or decreasing the temperature $T$. Furthermore, the analytic 
continuation of the data on the imaginary time or frequency 
axis to the real axis represents a difficult and 
numerically ill-conditioned problem
(see \cite{JG} for the application of the maximum entropy method 
to this problem).

Another non-perturbative method applicable here
is the exact diagonalization technique (see, e.g.,
\cite{Geo96,Caf94,Kra00}). 
In this method, the continuous conduction band of the effective
SIAM is approximated by a  discrete set of states 
(approximately 8-12 states). The value of $U$ does not impose a problem
here as the impurity (together with the conduction electron states)
is diagonalized exactly. The main disadvantage of the 
exact diagonalization technique is its inability to resolve
low energy features such as  a narrow quasiparticle
resonance at the Fermi level.

The above-mentioned restrictions concerning the
value of $U$ and $T$, or the low energy resolution,
do not apply to the numerical renormalization group
 method (NRG)\cite{Wil75,Kri80} 
which has only recently been used to investigate lattice
models within the DMFT \cite{Sak94,BHP,B,Pie99,Pru00}.
 The NRG as well has its drawbacks,
which will be discussed in Sec.~II of this paper; nevertheless
one would expect the NRG to be an ideal tool to calculate
the self-energy of the effective Anderson model in the
DMFT, simply because it has proven to be very successful
in the investigation of the physics of the standard SIAM.
For example, the NRG method is able to resolve, both
in static and dynamic properties,
the exponentially small Kondo scale for large values of $U$
(which can be seen neither in quantum Monte-Carlo nor in exact diagonalization).
One can also study in detail the scaling spectrum of the
quasiparticle peak \cite{Fro86,BGLP} and the temperature dependence of 
transport properties \cite{Cos92,Cos94} of the SIAM.

Applications of the NRG within the DMFT include
the investigation of the Mott-transition 
\cite{Sak94,BHP,B}, the problem
of charge ordering in the extended Hubbard model \cite{Pie99},
and the formation of the heavy-fermion liquid in the 
periodic Anderson model \cite{Pru00}. In all these investigations,
the temperature was restricted to $T\!=\!0$.

In this paper, we present a study of a strongly correlated
lattice model within the DMFT by applying the NRG method at
{\em finite} temperatures \cite{Cos92,Cos94}. In particular, 
we address a problem which has been the topic of an intense 
debate over the last couple of years: the details of the 
Mott-transition from a paramagnetic metal to a
paramagnetic insulator in the half-filled Hubbard model 
\cite{Geo96,B,BUCH,Sch99,Roz99,NG,Joo00}.

The paper is organized as follows: the NRG method is introduced
in Sec.~II, with particular emphasis on the calculation of finite
temperature dynamics.
In Sec.~III, the previous results for the Mott-transition
in the Hubbard model (within DMFT) are discussed.
The results from the NRG for the finite temperature Mott-transition
are then presented in Sec.~IV. The paper is concluded with 
a summary in Sec.~V.

\section{The Numerical Renormalization Group Method at finite
          temperatures\label{sec:2}}

\subsection{General Concepts}

The basic ideas of the NRG method were developed
by Wilson for the investigation of the Kondo model \cite{Wil75}.
Krishna-murthy,  Wilkins, and Wilson \cite{Kri80} 
later applied the NRG
to a related model, the single impurity Anderson model (SIAM)
with the Hamiltonian
\begin{eqnarray}
  H &=&   \sum_{\sigma} \varepsilon_{\rm f} f^\dagger_{\sigma}
                             f_{\sigma}
                 + U  f^\dagger_{\uparrow} f_{\uparrow}
                       f^\dagger_{\downarrow} f_{\downarrow}
                \nonumber \\
           & & + \sum_{k \sigma} \varepsilon_k c^\dagger_{k\sigma}
c_{k\sigma}
            +  \sum_{k \sigma} V
           \Big( f^\dagger_{\sigma} c_{k \sigma}
               +   c^\dagger_{k\sigma} f_{\sigma} \Big) \ . 
    \label{eq:siam}
\end{eqnarray}
In the model (\ref{eq:siam}), $c_{k\sigma}^{(\dagger)}$ denote 
annihilation
(creation) operators for band states with 
spin $\sigma$ and energy $\varepsilon_k$,
$f_{\sigma}^{(\dagger)}$
those for impurity states with spin $\sigma$ and energy 
$\varepsilon_{\rm f}$. The
Coulomb interaction for two electrons at the impurity site is given by
$U$ and
both subsystems are coupled via a hybridization $V$.

The hybridization function
\begin{equation}
  \Delta(\omega) = \sum_k \frac{V^2}{\omega - \varepsilon_k} \ ,
\end{equation}
is usually assumed to be constant between the band-edges
$-D$ and $D$, but will acquire some frequency dependence
in the effective Anderson model within the DMFT (the necessary
changes in the NRG-procedure due to the non-constant
$\Delta(\omega)$ were discussed in \cite{BHP,BPH}).

The first step to set up the renormalization group transformation 
is a logarithmic
discretization of the conduction band: 
the continuous
conduction band is divided into infinitely many intervals
$[\xi_{n+1},\xi_n]$ and $[-\xi_{n},-\xi_{n+1}]$ with 
$\xi_n = D \Lambda^{-n}$ and $n=0,1,2,\ldots,\infty$.
Here, $\Lambda$ is the
NRG discretization parameter (typical values used in the calculations
are $\Lambda = 1.5,\ldots,2$). The conduction band states in each 
interval are then replaced by a {\em single} state. While this 
approximation by a discrete set of states involves some coarse graining
at higher energies, it captures arbitrarily small energies near 
the Fermi level.

In a second step, this discrete model is mapped on a semi-infinite chain
form via a tridiagonalization procedure (for details, see \cite{Wil75,Kri80}
and section 4.2 in \cite{hew1}).
The Hamiltonian of the semi-infinite chain has the following form:
\begin{eqnarray}
  H &=& \sum_{\sigma} \varepsilon_{\rm f} f^\dagger_{\sigma}
                             f_{\sigma}
                 + U  f^\dagger_{\uparrow} f_{\uparrow}
                       f^\dagger_{\downarrow} f_{\downarrow}
                          + \sum_{\sigma} V \Big(
                f^\dagger_{\sigma} c_{0 \sigma}
              +  c^\dagger_{0 \sigma} f_{\sigma}  \Big)  \nonumber \\
&+& 
\sum_{\sigma, n=0}^\infty t_n \Big(
                c^\dagger_{n \sigma} c_{n+1 \sigma}
              +  c^\dagger_{n+1 \sigma} c_{n \sigma}  \Big) \ .
\label{eq:chain}
\end{eqnarray}
This form is valid for a general symmetric conduction band density of states.
The impurity now couples only to a single fermionic degree
of freedom (the $c^{(\dagger)}_{0 \sigma}$), with a hybridization
$V$.
Due to the logarithmic discretization, the hopping
matrix elements decrease as $t_n\propto\Lambda^{-n/2}$.
This means that, in going along the chain, the parameters in
the Hamiltonian evolve
from high energies (given by $D$ and $U$) to arbitrarily
low energies (given by $D\Lambda^{-n/2}$). The renormalization group
transformation is now set up in the following way.

We start with the solution of the isolated impurity, i.e., with
the knowledge of all eigenstates, eigenenergies, and matrix elements.
The first step of the renormalization group
transformation is to add the first conduction electron site, set up
the Hamiltonian matrices for the enlarged Hilbert space, and obtain
the new eigenstates, eigenenergies, and matrix elements
by diagonalizing these matrices. This procedure is then iterated.
An obvious problem occurs after only a few steps of 
the iteration. The Hilbert space grows as $4^N$ (with
$N$ the size of the cluster), which makes it impossible
to keep all the states in the calculation. Wilson therefore devised
a very simple truncation procedure in which only those
states with the lowest energies (typically
a few hundred) are kept. This truncation scheme
is very successful but relies on the fact that the hopping matrix
elements are falling off exponentially. High-energy states therefore
do not change the very low frequency behaviour and can be neglected.
This procedure gives for each cluster a set of eigenenergies and matrix
elements from which a number of physical properties can be derived.

\subsection{Finite Temperature Dynamics}

Here we want to discuss in detail the calculation of the
{\em finite} temperature spectral function
\begin{equation}
  A_\sigma(\omega) = -\frac{1}{\pi} {\rm Im}\ G_\sigma(\omega + i\delta^+) \ ,
\end{equation}
with
\begin{equation}
 G_\sigma(z) = i\int_0^\infty {\rm d} t \ 
e^{izt}\langle[f_\sigma(t),f_\sigma^\dagger]_+\rangle \ .
\end{equation}
From the iterative diagonalization described above, one can easily calculate
the spectral functions for each cluster of size $N$ via\cite{Walter}
\begin{eqnarray}
   A_{\sigma N}(\omega) &=& \frac{1}{Z_N}\sum_{nm}  
           \bigg\vert  
      \phantom{\Big\vert}_N  \Big< n \Big\vert f^\dagger_{\sigma}
                        \Big\vert m \Big> \!\!\phantom{\Big\vert}_N
                   \bigg\vert^2
                   \delta \big( \omega -(E^N_{n} -E^N_{m}) \big) 
         \nonumber \\
         & & 
  \times 
\left( e^{-\beta E^N_m} + e^{-\beta E^N_n} \right) \ .
    \label{eq:Ageneral}
\end{eqnarray}
Here $\big\{\big\vert n \big>_N\big\}$ 
and $\big\{\big\vert m \big>_N\big\}$ are sets of eigenstates
of the Hamiltonian for the cluster of size $N$, $E^N_{n}$ and  $E^N_{m}$
are the corresponding eigenenergies and $Z_N$ the
 grand canonical partition function (the spin index
$\sigma$ will be dropped in the following).
As the length of the cluster is successively increased, and
the $A_{\sigma N}(\omega)$ calculated in each step, Eq.~(\ref{eq:Ageneral})
defines a whole set of spectral functions. These data
are combined to give spectral functions as shown, e.g., in 
Fig.~\ref{fig:4.1} in the following way.

Let us first describe the procedure for calculating
the $T=0$ spectral function \cite{Cos92,Sak89}. The diagonalization
of the clusters $N=0,1,2,\dots$ yields the excitation
spectrum $\omega_{nm}=E_{n}^{N}-E_{m}^{N}$  on a set of decreasing energy 
scales $\omega_0 > \omega_1 > \omega_2 > \dots$ 
($\omega_N$ is the smallest scale in the truncated Hamiltonian $H_N$, i.e.,
$\omega_N=t_N$ and for a flat band one has 
$\omega_N\sim D\Lambda^{-(N-1)/2}$).
Excitations $\omega \ll \omega_N$ are not described  within
cluster $N$. They are obtained accurately in subsequent iterations from larger
clusters. Similarly, excitations  $\omega \gg \omega_N$ are outside
the energy window for cluster $N$ (whose width is limited on the high energy
side by the truncation of the spectrum). Information on these excitations
is contained in previous iterations for some smaller cluster $N'<N$. 
It is therefore possible to use Eq. (\ref{eq:Ageneral}) for each 
$N=0,1,\dots$ to calculate the $T=0$ spectral density at an appropriate
set of decreasing frequencies for each cluster. These frequencies
are chosen to be $\omega\approx 2\omega_N$ within the energy window of the
cluster under consideration (whose width, in units of $\omega_N$, 
typically lies in the range 6-10 for $\Lambda=1.5-2.0$).

\begin{figure}[ht]
\begin{center}\mbox{}
\psfig{figure=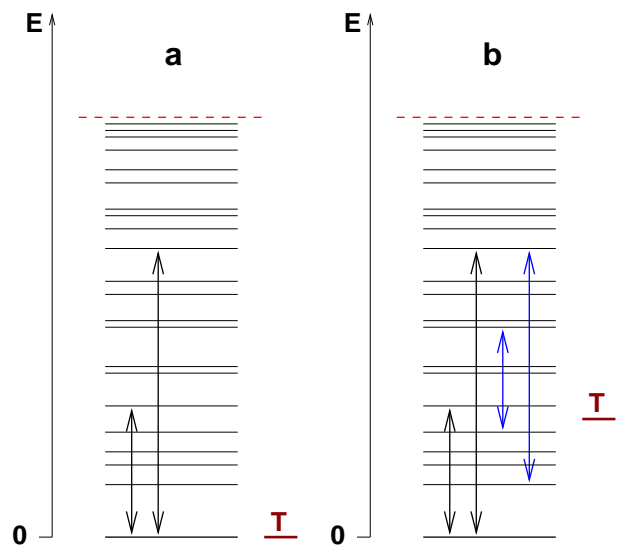,width=0.45\textwidth}
\end{center}
\caption[]{The spectrum of many-body excitations measured
with respect to the ground-state energy $E_{\rm g}\! =\! 0$,
and the possible transitions contributing to the single-particle
spectral function.~a) For $T\!=\!0$, only transitions between the
ground-state and excited states are possible; b) for $T\!>\!0$,
transitions between excited states are possible as well. The
dotted line indicates the cut-off in the spectrum due to the
truncation of states.}
\label{fig:2.1}
\end{figure}

At finite temperature, the above procedure is modified as follows. 
For a given temperature $T$, which we identify with 
$T_M\approx \omega_M$ for some $M$, one evaluates the 
spectral density in Eq. (\ref{eq:Ageneral}) at the
same characteristic frequencies $\omega=2\omega_{N}$ as those used
for the $T=0$ calculation, down to a minimum frequency corresponding to 
$\omega\approx T=T_M$. Compared to the $T=0$ calculation, many more 
excitations will contribute at finite $T$, as shown in 
Fig.~\ref{fig:2.1}. When $\omega=2\omega_N$ becomes comparable to 
or smaller than the temperature of interest, $T=T_M$, it is clear 
that excitations will start to contribute to the spectral density
at frequency $\omega$ which are not contained in cluster $N$. It
is still possible to calculate the spectral density at frequencies
$\omega=2\omega_N$ such that $-T_M \leq \omega \leq +T_M$ by using
the cluster of size $M$ corresponding to the temperature. 
This is achieved by broadening the $\delta$-functions in
the spectrum of cluster $M$ with broadening functions of width 
$T$ (see below).
This gives 
a very good estimate of the leading contribution to the spectral 
density for all frequencies $|\omega| \leq T$. It recovers, for
example, the known 
Fermi liquid relations for transport quantities of the Anderson model
\cite{Cos92,Cos94}.
Due to the limited resolution, proportional to 
$T_M$, the above scheme will, however, tend to broaden the
spectral densities too much at higher temperatures. 

This is not a problem
for the finite temperature spectral densities presented in this
paper. The reason, as we shall see below, is that the width of
the Kondo resonance in the effective impurity model is always
very much larger than the temperatures of interest (typically by a factor
of 10 larger).

The above scheme becomes increasingly
accurate as the temperature is lowered, eventually connecting continuously
the finite and zero temperature spectral densities as $T\rightarrow 0$.

There are several ways to put together the discrete information 
from the clusters in order to arrive at continuous curves for
spectral densities. One approach \cite{Cos92,Cos94} replaces
the delta functions in Eq.~(\ref{eq:Ageneral}) by appropriate
broadening functions 
(see Eq.~(\ref{eq:lorentzian}--\ref{eq:log-gaussian})) and evaluates
the spectral densities at the characteristic frequencies defined
above. It is also possible to first combine information on the
discrete spectra from successive clusters ($N$ and $N+2$,
to avoid even/odd effects) and
then broaden the spectra \cite{BHP}. Below, we describe this latter
approach, which we used to obtain most results in this paper.
A comparison between the two approaches gave only minor differences 
in the results for the spectral function.

\begin{figure}[ht]
\begin{center}\mbox{}
\psfig{figure=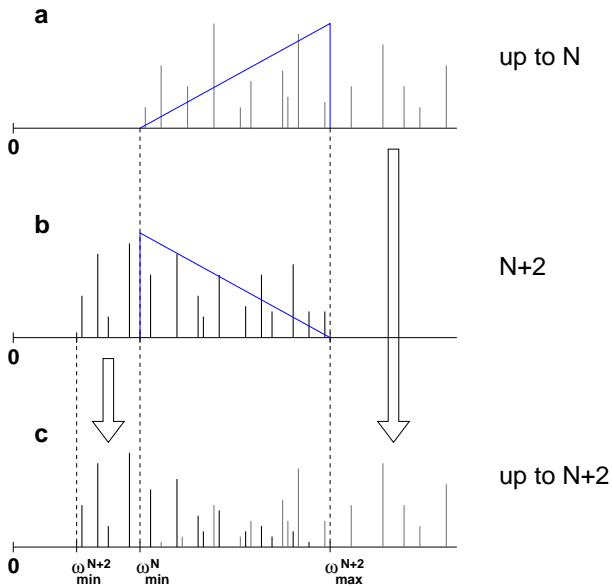,width=0.45\textwidth}
\end{center}
\caption[]{Superposition of the $\delta$-peaks in the spectral density
of all clusters up to length $N$ [see (a)] with the $\delta$-peaks
of the cluster of length $N+2$ [see (b)]. This procedure
gives the spectral information contained in all 
clusters up to length $N+2$ [see (c)]. The spikes indicate the weight of the
$\delta$-functions in the spectral density, and the lines in 
(a) and (b) correspond to the weighting function as described in the
text. The  $\delta$-peaks in the interval 
$[\omega_{\rm min}^{N+2},\omega_{\rm min}^N]$ and for 
$\omega > \omega_{\rm max}^{N+2}$ appearing in (c)
are identical to those appearing in (a) and (b), repectively,
as indicated by the arrows.}
\label{fig:2.2}
\end{figure}

The starting point is the set of $\delta$-peaks obtained
for a small cluster of size $N$ where the truncation
is not yet effective (see Fig.~\ref{fig:2.2}a).
The spectral 
distribution for the cluster of length $(N+2)$
is shown in Fig.~\ref{fig:2.2}b. 
The minimal frequency appearing in the spectrum for
the cluster of length $(N+2)$, $\omega_{\rm min}^{N+2}$,
is reduced approximately by a factor of $\Lambda$ 
compared to the frequency $\omega_{\rm min}^{N}$,
while
the  maximal frequency 
$\omega_{\rm max}^{N+2}$ is now determined by the number
of states retained after truncation. From the two sets
of $\delta$-peaks we keep those peaks which are
in the interval $[\omega_{\rm min}^{N+2},\omega_{\rm min}^N]$
and above $\omega_{\rm max}^{N+2}$. The peaks
in the overlapping region 
$[\omega_{\rm min}^N,\omega_{\rm max}^{N+2}]$  are 
taken from both the previous clusters and the one of length
$N+2$, and are added with a weighting function which is,
for simplicity, just a linear function with values 
from 0 to 1 for arguments between
$\omega_{\rm min}^N$ and $\omega_{\rm max}^{N+2}$ (for
the previous clusters) and with values from
1 to 0 (for the cluster of length $N+2$) \cite{wf}. 
The resulting set
of $\delta$-peaks is shown in Fig.~\ref{fig:2.2}c and 
can then be used to further iterate this procedure (with 
the cluster of length $N+4$, and so on), up to 
the cluster of length $M$ defined by $T=T_M$.

The resulting spectrum is still discrete.
To visualize the distribution of spectral
weight  it is convenient to broaden the $\delta$-peaks
using appropriate broadening functions. For the results
shown in this paper we used a Lorentzian 
\begin{equation}
  \delta(\omega - \omega_n) \rightarrow 
  \frac{1}{2\pi} \frac{b}{(\omega - \omega_n)^2 + b^2} \ ,
\label{eq:lorentzian}
\end{equation}
with width $b=0.6T$ for $\omega_n<4T$ and
a Gaussian on a logarithmic scale
\begin{equation}
  \delta(\omega - \omega_n) \rightarrow 
          \frac{e^{-b^2/4}}{b\, \omega_n\sqrt{\pi}} \exp \left[ 
      -\frac{(\ln\omega - \ln \omega_n)^2}{b^2} \right] \ ,
\label{eq:log-gaussian}
\end{equation}
with width $b=0.3$ for $\omega_n>4T$. 

So far, we have not made any reference to the application of the NRG 
to the effective Anderson model in the DMFT. The necessary steps
are described in \cite{BHP} for the case of $T\!=\!0$
and can be used for finite temperatures equally well. In particular,
the expression of the self-energy via
\begin{equation}
    \Sigma_\sigma(\omega) = U \frac{F_\sigma(\omega)}{G_\sigma(\omega)} ,
\label{eq:Sigma}
\end{equation}
with the correlation function 
$F_\sigma(\omega) = \langle\!\langle f_{\sigma} f^\dagger_{\bar{\sigma}}
f_{\bar{\sigma}},
      f^\dagger_{\sigma} \rangle\!\rangle_\omega $
holds for both $T\!=\!0$ and $T\!>\!0$ (for a discussion of the advantage
of using Eq.~(\ref{eq:Sigma}) for the calculation of $\Sigma(\omega)$,
see \cite{BHP}).

Let us now comment on the choice of temperatures used
in the calculations reported in this paper. It is clear from the
above discussion that the temperatures are chosen to
lie within the excitation spectrum of the cluster $M$
for which the NRG iteration is terminated. Keeping the
position of $T_M$ {\em within} the excitation spectrum
constant, one has to reduce $T_M$ by a factor
$\Lambda$ when the largest cluster is of length $M+2$.
This defines a discrete set of temperatures $T_N = T_M
\Lambda^{(N-M)/2}$ for which we perform the NRG calculations.

For a variety of applications within the DMFT one would certainly
prefer to vary the temperature {\em continuously} (to find, e.g.,
the critical temperatures for a phase transition).
Such a continuous variation is difficult within NRG.
It is certainly possible to achieve a large variation in
 temperature by a modest variation in $\Lambda$ and using
a fixed length of the cluster (due to the exponential
dependence of $T_M$ on $\Lambda$). The results obtained
in this way would, however, contain different systematic
errors, as the accuracy of the NRG is enhanced upon
reducing $\Lambda$. One should therefore try to correct
this $\Lambda$-dependence, e.g., by extrapolating the
results to $\Lambda=1$.
We have not attempted 
 to correct for the $\Lambda$-dependence and instead
worked with a fixed $\Lambda=1.64$ and different
cluster sizes (the number of states retained after truncation
is 600, not counting degeneracies).

\section{The Mott-Hubbard Metal-Insulator Transition\label{sec:3}}

Let us now turn to the 
 Mott metal-insulator transition \cite{BUCH,Mott}
 from a paramagnetic metal to
a paramagnetic insulator. This transition 
is found in various transition metal
oxides, such as $\rm V_2O_3$ doped with Cr \cite{McW}.
The mechanism driving the Mott-transition is believed to be the
local Coulomb repulsion $U$ between electrons on the same lattice site, although
the details of the transition should also be influenced by lattice degrees
of freedom. 
The simplest model to investigate the correlation driven metal-insulator
transition is the one-band Hubbard model \cite{Hub,Gut,Kan}
\begin{equation}
   H = -\sum_{ij\sigma} t_{ij}(c^\dagger_{i\sigma} c_{j\sigma} +
                   c^\dagger_{j\sigma} c_{i\sigma}) +
         U\sum_i c^\dagger_{i\uparrow} c_{i\uparrow}
            c^\dagger_{i\downarrow} c_{i\downarrow} \ ,
\label{eq:H}
\end{equation}
where $c^\dagger_{i\sigma}$ ($c_{i\sigma}$) denote creation
(annihilation) operators for a fermion on site $i$ and the $t_{ij}$ are the
hopping matrix elements between site $i$ and $j$ \cite{spin_one_half}.
Despite its simple structure, the solution of this model turns out to
be an extremely difficult many-body problem. The situation is particularly
complicated near the metal-insulator transition where $U$ and the
bandwidth are
roughly of the same order such that perturbative schemes (in $U$ or $t$)
are not applicable.

The existence of a metal-insulator transition in the 
paramagnetic phase \cite{frustration} of the half-filled Hubbard model has 
been known since the early work of Hubbard \cite{Hub}. The details of the
transition, however, remained unclear, except in the particular case
of dimension $d=1$, where the transition occurs at $U=0^+$
 \cite{BUCH,Lie68}. Even in the opposite limit of  infinite
dimensions, where a numerically exact solution of the Hubbard model is 
in principle possible, a general consensus concerning the 
details of the transition scenario
has not been reached so far.

Neglecting the transition to an antiferromagnetic phase
or suppressing it by frustration \cite{Geo96},
two coexisting solutions are found in DMFT at very low temperatures,
one insulating and one metallic \cite{ins-met}.
The transition is then of first order, even in the absence of a coupling
to lattice degrees of freedom. The scenario of a first-order transition
was first proposed in Refs. \cite{Geo93,Roz94}, within 
calculations based 
on the iterated perturbation theory and exact diagonalization.
It was later confirmed by the NRG for $T=0$ \cite{B}
and quantum Monte-Carlo 
calculations for $T>0$ \cite{Roz99,Nils}.
Criticism of this
scenario can be found in Refs.~\cite{NG,Kehrein,David,noz}.

The results from the NRG for the $T\!=\!0$ metal-insulator
transition can be summarized as follows (for
details see \cite{B}).
       On approaching the transition from the metallic side, a typical
       three-peak structure appears in the spectral function, with upper and
       lower Hubbard bands at $\omega \approx \pm U/2$ and a 
       quasiparticle peak at $\omega=0$. The width of the quasiparticle
       peak vanishes for $U\to U_{\rm c2}$, leaving behind two 
       well-separated Hubbard peaks 
       (see Fig.~2 in \cite{B}). 
       Although the NRG is not able to
       resolve a small spectral weight between the Hubbard peaks,
       the results indicate that the gap opens discontinuously
       (see also \cite{Geo96}).
       On decreasing $U$, the transition from the insulator to the
       metal occurs at a lower critical value $U_{\rm c1}$, where the gap
       vanishes.
Concerning the numerical value of $U_{\rm c2} \approx 1.47 W$
($W$: bandwidth)
excellent agreement with the result from the projective
self-consistent method \cite{Moeller,Geo96} 
is found.

The extension of the NRG to $T>0$ will now be used to
determine the full shape of the  hysteresis
region non-perturbatively.

\section{Results\label{sec:4}}

\subsection{Spectral function for $T>T_{\rm c}$}

\begin{figure}[t]
\begin{center}\mbox{}
\psfig{figure=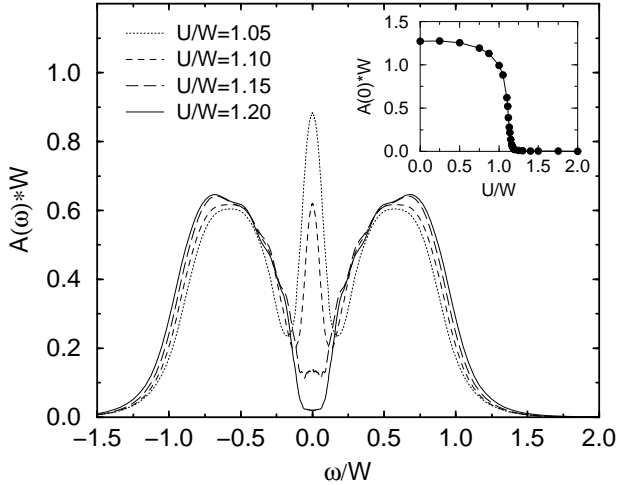,width=0.5\textwidth}
\end{center}
\caption[]{Spectral function for the half-filled Hubbard model for
various values of $U$ at $T=0.0276W > T_c$ (in the crossover region). 
The crossover from the
metal to the insulator occurs via a gradual suppression of
the quasiparticle peak at $\omega\!=\!0$. The inset shows
the $U$-dependence of $A(\omega\! =\!0)$, in particular
the rapid decrease for $U\approx 1.1W$.
}
\label{fig:4.1}
\end{figure}

Figure \ref{fig:4.1} shows the spectral function $A(\omega)$ for
various values of $U$ at $T=0.0276W$. This is above
the temperature of the critical point 
(which we estimate as $T_{\rm c}\approx 0.02 W$), 
so that
there is no real transition but a 
crossover from a  metallic-like to an insulating-like solution.
As already found in Refs.\cite{Geo96,Sch99},
the crossover region is nevertheless very narrow, with a very
rapid suppression of the quasiparticle peak. This is seen also
in the NRG results (Fig.~\ref{fig:4.1}) when 
$U$ is increased from $U=1.05W$ to $U=1.20W$. The spectral weight of the 
quasiparticle peak is gradually redistributed and shifted to the 
upper (lower) edge of the lower (upper) Hubbard
band. An additional structure {\em within} the Hubbard bands,
as reported in \cite{Geo96,Sch99}, is not found and would be very 
difficult to see due to the limited resolution of the NRG at higher
frequencies.

The inset of Fig.~\ref{fig:4.1} shows the $U$-dependence
of the value of the spectral function at zero frequency 
$A(\omega\!=\!0)$.
For higher values of $U$, the spectral density at the Fermi level
is still finite and vanishes only in the limit $U\to\infty$ (or for $T\to 0$,
provided that $U>U_{\rm c2}(T=0)$).

\begin{figure}[ht]
\begin{center}\mbox{}
\psfig{figure=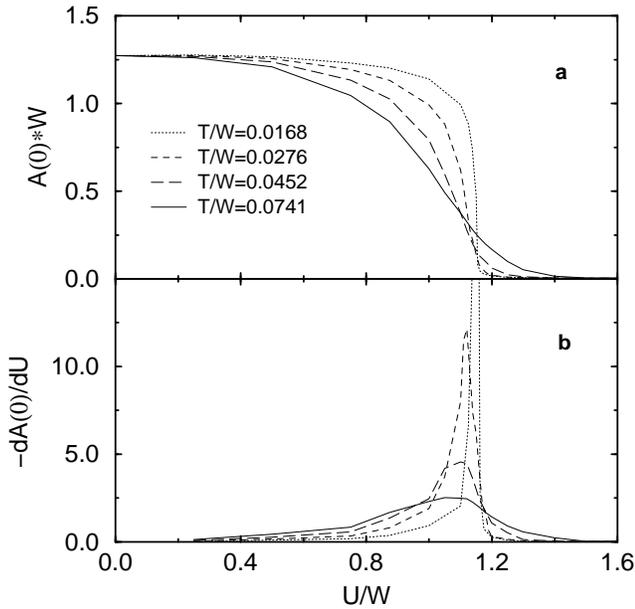,width=0.5\textwidth}
\end{center}
\caption[]{(a): The $U$-dependence of $A(\omega\! =\!0)$ for 
different temperatures; the data for $T\!=\!0.0168W<T_{\rm c}$ show
a very small hysteresis, not visible on this scale. The other 
three sets of data are for
$T> T_{\rm c}$. (b): The derivative
of  $A(\omega\! =\!0,U)$ with respect to $U$ for the same
temperatures as in (a).}
\label{fig:4.2}
\end{figure}

The $U$-dependence of $A(\omega\! =\!0)$ is shown
in Fig.~\ref{fig:4.2}a for different temperatures.
As discussed in Sec.~II,
the temperatures are chosen as
$ T_m = T_1 \cdot \Lambda^m $, with
$T_1 = 0.0168W$ and $m=0,1,2,3$,
($\Lambda = 1.64$ is used for all results shown in this
paper; the number of states retained after truncation
is 600, not counting degeneracies). 
The derivative of $A(\omega\! =\!0)$ with respect to $U$,
\begin{equation}
  A^\prime(\omega\! =\!0) \equiv 
        \frac{\partial A(\omega =0,U)}{\partial U} \ ,
\end{equation}
is plotted in Fig.~\ref{fig:4.2}b. 
The $U$-value where $ \vert A^\prime(\omega\! =\!0) \vert$ 
reaches its maximum defines a characteristic interaction
strength $U^\ast$ for the crossover from metallic-like
to insulating-like behavior in the region $T\!>\!T_{\rm c}$;
for the definition of  $U_{c1,2}$ for
$T\!<\!T_{\rm c}$, see below. 
Furthermore, the width $\Delta U$ of the crossover region
 can be defined as the width at half-height of the
peak in $ A^\prime(\omega\! =\!0) $.

Upon lowering the temperature, the width $\Delta U$ rapidly decreases
and vanishes at the critical temperature $T_{\rm c}$, since 
$A^\prime(\omega\! =\!0)$ diverges at $T_{\rm c}$ (this feature
has already been discussed in \cite{Kot00}).
A precise value for $T_{\rm c}$ cannot be given
 as we are presently not able to  vary
the temperature within NRG continuously.
The critical temperature is estimated as $T_{\rm c}\approx 0.02 W$,
as a very small hysteresis is still present for $T=0.0168W$ (on the
scale of Fig.~\ref{fig:4.2}, $U_{\rm c1}$ and $U_{\rm c2}$ cannot
be distinguished).

The $U^\ast$ as defined above slowly decreases upon increasing
the temperature.
This is not  at variance with the opposite
trend observed in Refs.
\cite{Geo96,Pru93,Sch99} (in Ref. \cite{Sch99}, the slope of
$U^\ast$ changes sign at $T\approx 0.25W$)
and depends on the definition of $U^\ast$.
Taking
$U^\ast$ as, e.g., the value of $U$ where $ A(\omega\! =\!0) $
has dropped to 1\% of its value at $U\!=\!0$ would result in an
{\em increase} of $U^\ast$ upon increasing the 
temperature. 

\subsection{Breakdown of Fermi Liquid versus Metal-Insulator Transition}

We now discuss the question of how to define a useful criterion for the
metal-insulator transition at finite temperatures. At zero
temperature a suitable criterion is the vanishing, with increasing $U$, 
of the quasiparticle weight 
\begin{equation}
  Z = \frac{1}{1-
      \left. \frac{\partial {\rm Re} \Sigma(\omega)}{\partial\omega}
      \right\vert_{\omega=0}} \ .
\label{eq:qpweight}
\end{equation}
The physical meaning of $Z$ is clear for the paramagnetic state at 
$T=0$, where the system is either a Fermi liquid (for $U<U_{\rm c}$) or 
an insulator (for $U>U_{\rm c}$). The vanishing of $Z$ therefore marks 
the metal-insulator transition at $T=0$, as discussed, e.g, in \cite{Geo96,B}. 
This criterion, however, cannot be taken over straightforwardly to
finite temperatures, since for $T>0$ the breakdown of the Fermi liquid
state and the appearence of the insulating state do not coincide.
Although this point has been noted before in the literature (see, e.g.,
\cite{Roz95,Sch99}), the vanishing of $Z$ has been used
as (one) criterion for the occurence of the metal-insulator transition
in \cite{Sch99}. It should be noted that the definition of $Z$
used in the finite temperature quantum Monte-Carlo calculations
of Ref. \cite{Sch99} is different from Eq. (\ref{eq:qpweight})
since  $\left. \partial {\rm Re} \Sigma(\omega)/\partial\omega
      \right\vert_{\omega=0}$
was approximated by ${\rm Im} \Sigma(i\omega_0)/\omega_0$,
with $\omega_0$ the first Matsubara frequency.

\begin{figure}[ht]
\begin{center}\mbox{}
\psfig{figure=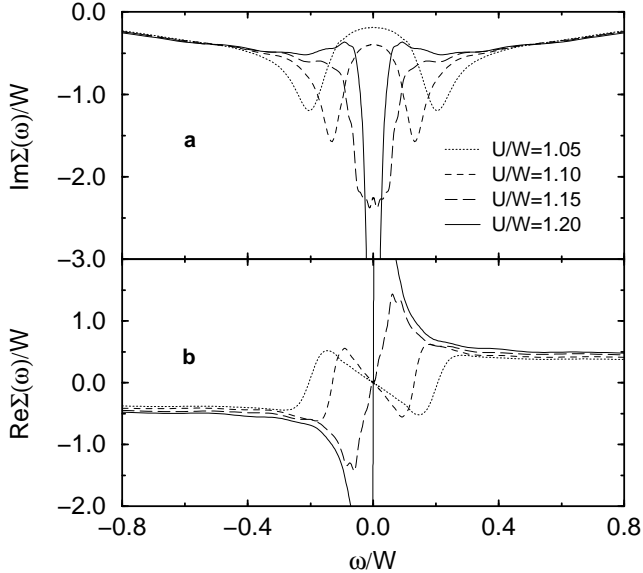,width=0.5\textwidth}
\end{center}
\caption[]{Imaginary part (a) and real part (b) of the self-energy for
the same temperature ($T=0.0276W$)
and $U$-values as
in Fig.~\ref{fig:4.1}. The slope of  ${\rm Re}\Sigma(\omega)$
changes sign at the same $U$-value for which the peak at
$\omega\!=\!0$ appears in Im$\Sigma(\omega)$.
}
\label{fig:4.3}
\end{figure}

To elucidate this point, it is instructive to discuss the behavior of 
the self-energy
in the crossover region from the metallic-like to the insulating-like solution.
The real and imaginary part of $\Sigma(\omega)$ are shown in 
Fig.~\ref{fig:4.3}, for the same temperature and $U$ values as
in Fig.~\ref{fig:4.1}. For $U=1.05W$ and $U=1.10W$ the imaginary part
shows the characteristic structure of the self-energy for a Fermi liquid 
(with the $\omega^2$ dependence for small frequencies and the
falling off at higher frequencies which leads to a two-peak structure),
but with a rapidly increasing scattering rate
at $\omega\!=\!0$ for increasing $U$. The two-peak structure
gradually evolves into a structure with a well-pronounced
peak at $\omega\!=\!0$  characteristic
for an insulating solution (a vanishing $ A(\omega\! =\!0) $
would correspond to a $\delta$-function in ${\rm Im}\Sigma(\omega)$).
Note that for $U\ge 1.15W$
 the value of ${\rm Im}\Sigma(\omega=0)$  is much larger than the
$T^2$-term observed for $T\to 0$. 
Hence, the mechanism
for the strong scattering at $\omega\! =\!0$ is {\em not} the 
quasiparticle-interaction    
but is caused by the bare local Coulomb repulsion, which 
destroys the Fermi liquid behavior for $U\ge 1.15W$.
 For $U=1.15W$ there is still a narrow dip in
${\rm Im}\Sigma(\omega)$ at $\omega\!=\!0$ corresponding 
to the remnant of the quasiparticle peak seen in Fig.~\ref{fig:4.1}.

For $U=1.05W$ and $U=1.10W$, the corresponding real part
of $\Sigma(\omega)$ shows the typical Fermi liquid behavior
with a negative slope at $\omega\! =\!0$. Upon further increasing
the $U$, however, the slope of  ${\rm Re}\Sigma(\omega)$
changes sign right at the $U$-value where the peak at
$\omega\!=\!0$ appears in Im$\Sigma(\omega)$;
this is obvious from the Kramers-Kronig transformation
which connects real and imaginary part. Note that
the $1/\omega$-behavior in ${\rm Re}\Sigma(\omega)$ for larger
frequencies is not visible on this scale.

\begin{figure}[ht]
\begin{center}\mbox{}
\psfig{figure=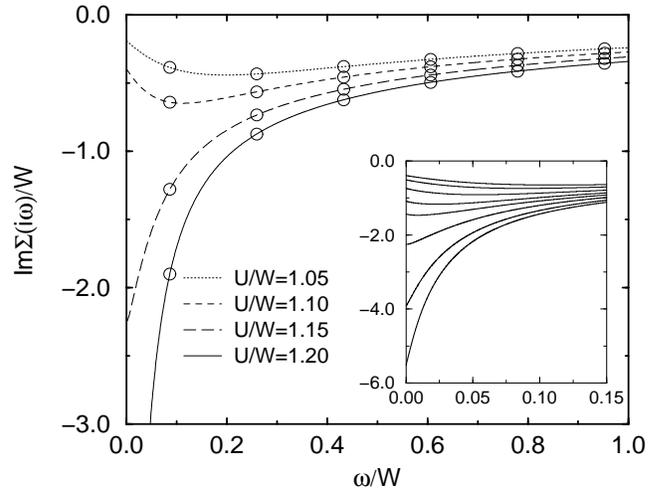,width=0.5\textwidth}
\end{center}
\caption[]{Imaginary part of the self-energy, Im$\Sigma$,
on the imaginary
frequency axis for the same parameters as in  Figs.~\ref{fig:4.1}
and \ref{fig:4.3}. The values of Im$\Sigma$  for the
Matsubara frequencies are indicated by the circles.
The inset focuses on the change of sign of the slope in
Im$\Sigma(i\omega)$ for values of $U=1.1W$ up to $U=1.17W$ (from
top to bottom).}
\label{fig:4.4}
\end{figure}

From the full frequency dependence of $\Sigma(\omega)$ on the real
axis one can easily perform the analytic continuation  to $\Sigma(z)$
for any value of $z$ in the upper complex plane:
\begin{equation}
  \Sigma(z) = -\frac{1}{\pi} \int {\rm d} \omega^\prime
        \frac{{\rm Im}\Sigma( \omega^\prime)}{z-\omega^\prime} \ .
\label{eq:an_con}
\end{equation}
In particular for $z=i\omega$ and $\omega$ real, Eq.~(\ref{eq:an_con})
gives the real and imaginary parts of the self-energy on the 
imaginary frequency
axis (the analytic continuation from the imaginary to the real frequency
axis, however, is much more delicate, see, e.g., 
\cite{JG}).
The result for Im$\Sigma(i\omega)$ is shown in Fig.~\ref{fig:4.4}, 
for the same parameters as in  Figs.~\ref{fig:4.1} and \ref{fig:4.3}.
The circles indicate the value of  Im$\Sigma(i\omega_n)$
for the Matsubara frequencies
\begin{equation}
   \omega_n = \frac{\pi}{\beta} (2n+1) , \ \ \ \ n=0,1,2,\ldots \ .
\end{equation}
As the self-energy $\Sigma(z)$ is defined on the whole imaginary
frequency axis, not only for the Matsubara frequencies, one can,
for instance, check the trivial condition 
${\rm Im}\Sigma(i\omega) = {\rm Im}\Sigma(\omega)$ for 
$\omega\to 0$.

Furthermore, the slope of ${\rm Im}\Sigma(i\omega)$
for $\omega\to 0$ is identical to the slope of the real part
of $\Sigma(\omega)$. As a consequence, the same change in
the slope of the self-energy 
is visible in  both Figs.~\ref{fig:4.3}b  and  \ref{fig:4.4}.
The inset of Fig.~\ref{fig:4.4} illustrates this for a smaller frequency
range and a narrow mesh of $U$-values from $U=1.1W$ up to $U=1.17W$ 
(from top to bottom).

This behavior of the self-energy has
drastic consequences for the notion of a quasiparticle weight $Z$
in the crossover regime. We see that the
application of Eq.~(\ref{eq:qpweight}) to the self-energies as obtained
in Figs.~\ref{fig:4.3} and \ref{fig:4.4} leads to unphysical results for
$U\ge 1.15W$.
Due to the change of sign in 
$\left. \frac{\partial {\rm Re} \Sigma(\omega)}{\partial\omega}
      \right\vert_{\omega=0}$ upon increasing $U$, the $Z$ from 
 Eq.~(\ref{eq:qpweight})  starts  increasing again and even diverges
at a particular value of $U$ for which the derivative of the self-energy
is equal to one. For larger values of $U$, $Z$ becomes negative and
approaches zero from below for $U\to \infty$. Apparently,
the use of Eq.~(\ref{eq:qpweight}) does not make sense for
$U\ge 1.15W$ \cite{U}
which is due to the fact the the concept of quasiparticles itself
breaks down in the crossover regime. The quasiparticle weight is therefore
not an appropriate measure for the metal-insulator transition 
in the whole parameter space.
Note also that in the crossover region,
the weight of the remnant of the quasiparticle peak
is not associated to $Z$.

Whereas there is no unique criterion for a characteristic value
$U^\ast$ for $T>T_{\rm c}$,
critical values for $U$ can nevertheless
be defined for $T<T_{\rm c}$ via the 
 value of $U$ at which $A(\omega=0)$ changes 
discontinuously. 

\subsection{Spectral function for $T<T_{\rm c}$}

\begin{figure}[ht]
\begin{center}\mbox{}
\psfig{figure=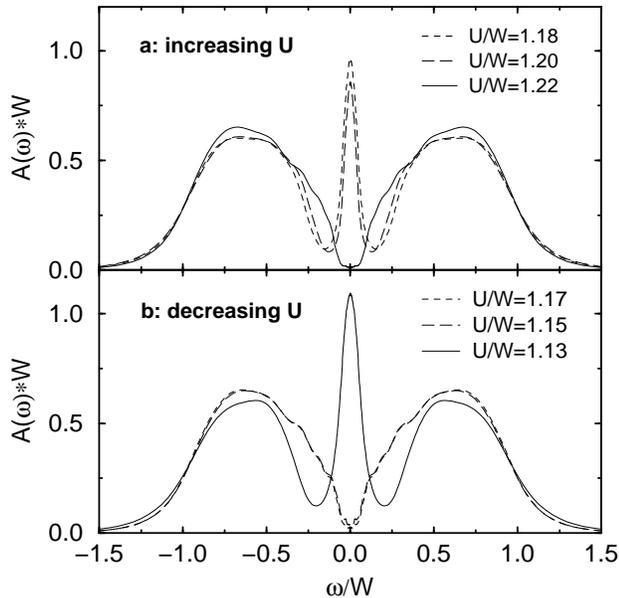,width=0.5\textwidth}
\end{center}
\caption[]{Spectral function for $T\!=\!0.0103W$; (a): increasing
$U$ (b): decreasing $U$; the transitions at $U_{\rm c2}\approx 1.21W$
and $U_{\rm c1}\approx 1.14W$ are characterized by a significant
redistribution of spectral weight and a jump in $A(\omega\!=\!0)$
(see also Fig.~\ref{fig:4.6}) .}
\label{fig:4.5}
\end{figure}

Figure \ref{fig:4.5} shows the spectral function $A(\omega)$ in the 
hysteresis region for $T\!=\!0.0103W$, both for increasing 
$U$ (Fig.~\ref{fig:4.5}a) and decreasing $U$ (Fig.~\ref{fig:4.5}b). 
 The results are shown for a very fine mesh of 
$U$-values close to $U_{\rm c2}\approx 1.21W$ 
and $U_{\rm c1}\approx 1.14W$.

\begin{figure}[ht]
\begin{center}\mbox{}
\psfig{figure=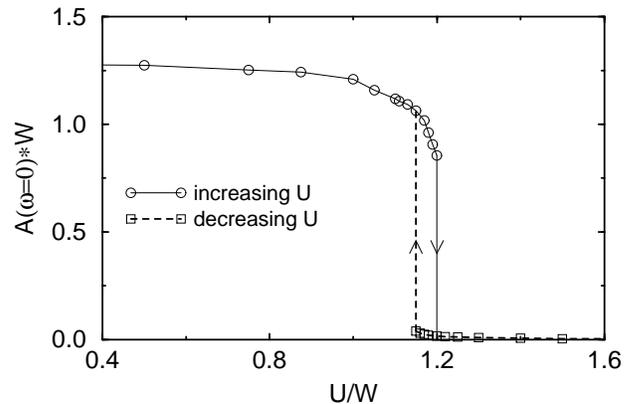,width=0.5\textwidth}
\end{center}
\caption[]{$U$-dependence of $A(\omega\!=\!0)$ for $T\!=\!0.0103W$;
solid line: increasing $U$, dashed line: decreasing $U$.}
\label{fig:4.6}
\end{figure}

In both cases, the transition is of first order, i.e., associated with a
discontinuous redistribution of spectral weight. 
The hysteresis effect is further
illustrated in the $U$-dependence of $A(\omega\!=\!0)$ for
$T\!=\!0.0103W<T_c$ (Fig.~\ref{fig:4.6}).

Whereas the critical values  $U_{\rm c1}$ and $U_{\rm c2}$ can
be easily defined by the jump in $A(\omega\!=\!0)$, the calculation
of the actual thermodynamic transition requires the knowledge of the free
energy $F$ of both metallic and insulating solutions. The
determination of $F$ goes beyond the scope of this paper. There
is no way of directly calculating $F$ within the NRG approach,
so one has to determine the free energy
 via integrating over a path from a particular point in
the $(U,T)$-plane for which the free energy is known, up
to the actual values of $U$ and $T$. However, the knowledge
of the $U_{\rm c}(T)$ for the actual thermodynamic transition
will not alter the fact that
the transition is of first order.

\subsection{Phase diagram}

Let us finally discuss the phase diagram for the Mott
metal-insulator transition in the very low temperature region.

\begin{figure}[ht]
\begin{center}\mbox{}
\psfig{figure=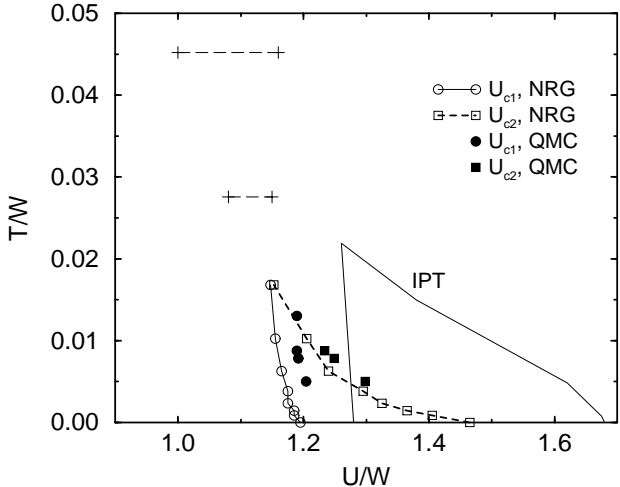,width=0.5\textwidth}
\end{center}
\caption[]{Results for the phase diagram of the Mott transition
obtained from different methods: NRG (open symbols), quantum Monte-Carlo
(QMC, filled symbols) and iterated perturbation theory (IPT, 
solid lines). The dashed lines for $T>T_{\rm c}$
indicate the position and width of the crossover region
as calculated from the data of Fig.~\ref{fig:4.2}.}
\label{fig:4.7}
\end{figure}

In Fig.~\ref{fig:4.7}, the dashed lines for $T>T_{\rm c}$
indicate the position and width of the crossover region
as calculated from the NRG data of Fig.~\ref{fig:4.2}.
The open circles  and squares  are the NRG-results
for $U_{\rm c1}(T)$ and $U_{\rm c2}(T)$, respectively. 
As the NRG calculations
cannot, so far, be performed for arbitrary values of $T$, we
cannot  give a precise value for the critical point. 
The $U_{\rm c2}(T)$ nicely
extrapolates to the previously obtained 
value for $T\!=\!0$ \cite{B}; the same is true for $U_{\rm c1}(T)$.
Note that the value for $U_{\rm c1}(T\!=\!0)= 1.195W$ plotted here
is slighly reduced as compared to the originally published value 
$U_{\rm c1}(T\!=\!0)= 1.25 W$ \cite{B}. This is due to the different
value for $\Lambda$, the number of states and the broadening
used here.

Figure \ref{fig:4.7} also contains recent quantum Monte-Carlo
results of Joo and Oudovenko \cite{Joo00} (filled symbols),
as well as the result from the iterated perturbation theory
\cite{Geo96} which tends to overestimate both $U_{\rm c1}(T)$
and $U_{\rm c2}(T)$.
The phase boundaries obtained from the NRG are 
below the values obtained from
the quantum Monte-Carlo
results. 
Concerning the NRG values, it is well-known that due to the
logarithmic discretization the NRG tends to underestimate the
effective hybridization 
\cite{Gon96} 
(hence underestimating the value of $U$ 
necessary to overcome the kinetic energy). This effect has, e.g.,
been studied in the context of the quantum phase-transition
from the local moment to the strong coupling phase in the soft-gap 
Anderson model \cite{BGLP}. For the transition in this model, the value
of $U_{\rm c}$ for $\Lambda\!=\!2.0$ is about 5\% below
the extrapolated value for $\Lambda\to 1$; more importantly,
the $U_{\rm c}(\Lambda)$ is a perfectly straight line
from $\Lambda\!=\!1.4$ to $\Lambda\!=\!3.0$.

A similar $\Lambda\to 1$ extrapolation is difficult to
perform for the metal-insulator transition studied here, since
already a very large number of DMFT iterations is necessary
to determine a single value of $U_{\rm c}(\Lambda)$. Calculations
of $U_{\rm c1}$ and $U_{\rm c2}$ for one value of
$T$ with
$\Lambda\!=\!1.64$ and $\Lambda\!=\!2.0$   at least show
the expected trend, i.e., a slight increase of the $U_{\rm c}$'s
with decreasing $\Lambda$.

Taking into account the unavoidable
numerical errors in both procedures, the agreement
between NRG and quantum Monte-Carlo results for
the phase-boundary is seen to be very good; the
agreement can even be further improved \cite{Nils}.

\section{Summary\label{sec:5}}

In this paper we presented results from the numerical
renormalization group method (NRG) for the finite
temperature Mott transition in the Hubbard model
on a Bethe lattice within dynamical mean field theory.
For the crossover region $T>T_{\rm c}$, the quasiparticle
peak in the spectral function gradually vanishes
upon increasing $U$ and the imaginary part of the self-energy
develops a sharp peak at $\omega=0$. Associated with this
is a change of sign of Re$\Sigma(\omega)$ at $\omega=0$.
As a consequence, the behaviour of the quasiparticle weight
can no longer be used as a criterion
for the transition at finite temperature. 

For $T<T_{\rm c}\approx 0.02W$, we find two coexisting solutions
in the range $U_{\rm c1}(T) < U < U_{\rm c2}(T)$. The values for the
critical $U$ can be determined for arbitrarily small temperatures,
in contrast to the quantum Monte-Carlo method which is so far
restricted to $T>W/150$. The critical $U_{\rm c1}(T)$ and
$U_{\rm c2}(T)$ are characterized by a redistribution of finite
spectral weight in the spectral function.

We therefore obtain a consistent picture for the
Mott metal-insulator transition from a paramagnetic metal
to a paramagnetic insulator in the whole parameter regime. 
The results are in very good
agreement with those from other non-perturbative methods
(the quantum Monte-Carlo method and the projective
self-consistent method) in their respective ranges of applicability.

There are still several questions left for further investigations.
A continuous variation of the temperature within the NRG
requires a better understanding of the $\Lambda$-dependence
of the results. 
The NRG also allows the calculation of a variety of
dynamic and transport properties in the
whole parameter regime, such as dynamic susceptibility
and optical conductivity. A generalization of the NRG method
to antiferromagnetic phases and the Hubbard model away
from half-filling is in progress.

\noindent{\bf Acknowledgements:} It is a pleasure to acknowledge fruitful
discussions with N. Bl\"umer, W. Hofstetter, 
A.P. Kampf, and Th. Pruschke.
Two of us (RB and TAC)  would like to thank the Isaac Newton Institute for 
Mathematical Sciences for hospitality where part of this work was done.
We also acknowledge the support of the Deutsche 
Forschungsgemeinschaft through the Sonderforschungsbereich 484.


\begin{thebibliography}{99}
\bibitem{wf} The precise form of the weighting function turns
              out to be unimportant for the final result, e.g., the
              form of the spectral function; the idea is
              here to reduce
              the weight of those $\delta$-peaks which are at the
              edges of the spectrum of each cluster.
\bibitem{frustration} The hopping matrix elements $t_{ij}$ in 
          Eq.~(\ref{eq:H}) are assumed to be chosen in such a way that
          antiferromagnetic long-range order is suppressed completely
          \cite{Geo96}.
\bibitem{ins-met} Although the term insulator is only strictly 
        defined for $T=0$, we denote
        a solution as insulating if it shows well separated upper
        and lower Hubbard peaks with a only a small spectral weight at the
        Fermi level.
\end{thebibliography}
\end{document}